\documentclass[twocolumn,prl,showpacs]{revtex4}

\usepackage{amsmath}
\usepackage{amssymb}
\usepackage{graphicx}

\makeatletter
\newcommand{\Ab}{\textbf{A} }		
\newcommand{\Bb}{\textbf{B} }
\newcommand{\Eb}{\textbf{E} }
\newcommand{\Fb}{\textbf{F} }
\newcommand{\Jb}{\textbf{J} }
\newcommand{\Pb}{\textbf{P} }
\newcommand{\pb}{\textbf{p} }

\newcommand{\sgn}{\ensuremath{\operatorname{sgn}}}

\begin{document}

\title{A non-linear theory for the bubble regime of plasma wake fields in tailored plasma channels}

\author{Johannes Thomas$^{1}$}\email{thomas@tp1.uni-duesseldorf.de}

\author{Igor Yu. Kostyukov$^{2,3}$}

\author{Jari Pronold$^{1}$}

\author{Anton Golovanov$^{2,3}$}

\author{Alexander Pukhov$^{1}$}

\affiliation{$^{1}${\small{}Institut f\"ur Theoretische Physik I, Heinrich-Heine-Universit\"at D\"usseldorf, D-40225 Germany}}

\affiliation{$^{2}${\small{}Lobachevsky State University of Nizhni
		Novgorod, 603950 Nizhny Novgorod, Russia}}

\affiliation{$^{3}${\small{}Institute of Applied Physics RAS, 603950 Nizhny Novgorod, Russia}}


\begin{abstract}
We introduce a first full analytical bubble and blow-out model for a radially inhomogeneous plasma in a quasi-static approximation. For both cases we calculate the accelerating and the focusing fields. In our model we also assume a thin electron layer that surrounds the wake field and calculate the field configuration within. Our theory holds for arbitrary radial density profiles and reduces to known models in the limit of a homogeneous plasma.
From a previous study of hollow plasma channels with smooth boundaries for laser-driven electron acceleration in the bubble regime we know that pancake-like laser pulses lead to highest electron energies [\textit{Pukhov et al, PRL 113, 245003 (2014)}]. As it was shown, the bubble fields can be adjusted to balance the laser depletion and dephasing lengths by varying the plasma density profile inside a deep channel. Now we show why the radial fields in the vacuum part of a channel become defocussing.
\end{abstract}
\maketitle

\subsection{Introduction}
Known analytical and semi-analytical models for broken plasma wave wake fields were introduced so far for homogeneous plasmas \cite{Lu2006, Lu2007, Kostyukov2004, Kostyukov2009, Kostyukov2010, Thomas2014, Kalmykov2011, Kalmykov2012, Yi2011, Yi2013, Pak2010, Zeng2012}. These models and extensive numerical simulations have shown that plasma wake fields provide a feasible path for high gradient particle acceleration \cite{Joshi2006,Esarey2009,Joshi2013}. Especially efficient are the so called bubble regime of laser-plasma wake fields \cite{Pukhov2002} and the blow-out regime of particle wake field acceleration \cite{Lu2006}.  For the laser driven case, the pulse is shorter than the plasma wavelength and fits perfectly into the first half of the plasma period. The laser intensity is thought to be high enough that the created wake field breaks after its first oscillation. In this regime, the wake field takes the form of a distorted spherical cavity from which all electrons are expelled and that moves with nearly the speed of light through the plasma. In the following we refer to this wake-field as "the bubble". In the case of a dense highly energetic particle bunch the blow-out regime is reached if the bunch density $n_b$ is larger than the unperturbed ambient plasma density $n_0$ and if both the axial and radial bunch lengths are short ($k_p\sigma_r<1$, $k_p\sigma_z<1$) \cite{Rosenzweig1991}. In both cases all plasma electrons are expelled from the region behind the driver and the driver creates a cavitated region that has a transversely uniform accelerating field \cite{Kostyukov2004}. This field helps to generate quasi-monoenergetic electron bunches readily registered in experiments \cite{Malka2013}. Despite various theoretical approaches to the basic analysis (the phenomenological model of the bubble \cite{Kostyukov2009,Kostyukov2010}, the nonlinear theory of blowout regime \cite{Lu2007, Lu2006, Lu2006a}, and the similarity theory \cite{Gordienko2004}) a self-consistent theoretical description of these regimes is still absent. 

Another aspect of the blow-out and the bubble regime is the self-injection physics that strongly depends on the fields near the ion cavity border. In this context a recent work of Yi et al. \cite{Yi2013} introduces an analytic model of the electric and magnetic fields surrounding the electron void in a plasma wake field accelerator. The model discusses the electron sheath and the resulting global fields inside and outside the void. Afterwards, the fields are used to describe electron self-injection in a plasma with a smooth density gradient. Another injection model that takes the action of the beam load into account and gives analytical solutions for the fields and the shape of the ion channel is presented in \cite{Tzoufras2008, Tzoufras2009}. Both models give an advance in the field of analytical modeling of trapping and injection but still hold for homogeneous background plasmas solely. Furthermore, the key to produce mono-energetic electron beams in the bubble or blow-out regime is to find the right driver configuration. Thus, for the bubble regime, scaling laws have been derived and extensively tested in 3d PIC simulations \cite{Gordienko2005, Pukhov2006, Jansen2014}. However, all this work has been done for a homogeneous background plasma so far. 

Former discussions of non-homogeneous plasmas in the context of electron acceleration target the guiding of relativistic laser pulses and their diffraction \cite{Sprangle1992,Geddes2004,Leemans2006}. Recently Schroeder et al. \cite{Schroeder2013,Schroeder2013a} suggested to use nearly hollow plasma channels to provide independent control over the focusing and accelerating forces in the cavity, Pukhov et al. \cite{Pukhov2014b} derived new scaling laws for deep plasma channels in which the density drops to zero on axis. Different to the idea to use plasma channels to guide weakly relativistic pulses over longer distances, Pukhov et al. use the channel to adjust the depletion length to the dephasing length and thus to enhance the energy gain. In this context the key role of the channel is to introduce new degrees of freedom that help to adjust measurable quantities as the energy gain, the bunch energy spread, and the trapping ratio. Due to this progress, analytical models for broken wake fields excited by plasma channel guided laser or particle drivers are required.

In this context the following work gives a first analytical model for fields inside an ion cavity in a plasma channel with an arbitrary radial profile. The basic idea for the model is a generalization of Lu's model \cite{Lu2007, Lu2006, Lu2006a} for a plasma channel. Different to \cite{Lu2006} we also calculate the fields inside the surrounding electron layer. However, the electron layer will be treated as a thin sheath with a thickness that is small compared to the radius of the bubble. In general our model self-consistently treats the nonlinear evolution of the driver but for reasons of simplicity all fields are calculated in a quasi-static theory. This implies that the electron cavity slowly evolves in time and that the fields depend on $\xi=ct-z$ solely \cite{Sprangle1990, Sprangle1990a, Whittum1997, Mora1997}. In our model the coordinates are normalized to the inverse electron wave number $k_{p}^{-1}=c/\omega_{p}$ while the time is normalized to $\omega_{p}^{-1}$. Here $\omega_{p}^{2}=4\pi e^{2}n_0/m_e$ is the electron plasma frequency and $n_0$ is a certain density in the system to which both $n_e(r)$ and the ion density $\rho_{ion}(r)$ are normalized. For example it is convenient to assume that the deep plasma channel is embedded into a homogeneous plasma. Then $n_0$ could be the unperturbed ambient plasma density or the density at a certain distance to the symmetry axis. The cavity surrounding electron layer has thickness $\Delta\ll \max(r_b)$ where $r_b(\xi)$ is the inner cavity radius. The electron layer screens the large inner potentials at the cavity border so that the outer potentials vanish in the unperturbed ambient plasma. The impact of the sheath thickness on the inner fields is neglected. The specific shape of the electron layer strongly depends on the radial plasma profile. As we show, the so far assumed spherical form in the analytical models for homogeneous plasmas is still valid in a region where the layer distance to the channel axis is large and where the influence of the driver is negligible. In general a reduction of the plasma density - i.e. as in a channel - leads to a decrease of the wake field length and a simultaneous steepening at the rear and front sides. Especially for plasma densities of the form $\rho_{ion}=\alpha r^n$ we are able to give a closed analytical expression for the cavity length. Our cylinder symmetric model recovers all known features of former models in the special case of uniform plasma $\rho_{ion}=1$.
\\

In the following sections we first introduce the model for an electron bunch-driven blow-out in an arbitrarily formed plasma channel. The model distinguishes between regions of high electron density and actually electron-free spaces inside the cavity. These regions are defined in Fig.\ref{model} where region 0 (gray) is free from electrons. They are located inside the electron sheath. Region I (red) is just inside the driving electron bunch and can be used to describe the back action of the sheath onto the driver and possible self-modulation. Region II (light red) is radially outside the driving electron bunch. Here the driver current affects both the magnetic field and the radial electric field. At last, region III (black dots) describes the fields inside the accelerated electron bunch while region IV (light blue) is radially outside the accelerated electron bunch. The explicit field configuration inside the electron layer (black) is calculated later.

In section \ref{sheath} we calculate the electron layer shape from the equations of motion for a test electron at the inner boundary and show why a plasma channel leads to longitudinal squeezing of the cavity. In section \ref{field} we derive the inner fields in terms of the radial blow-out radius $r_b$ and the ion density profile. Afterward, we give the sources in terms of the fields and the radius $r_b$. In section \ref{reverse} we show how the additional freedom provided by the channel can be used to reverse the radial electrical field.

\subsection{The bubble model in a plasma channel}
\begin{figure}[t]
	\includegraphics[width=1\columnwidth]{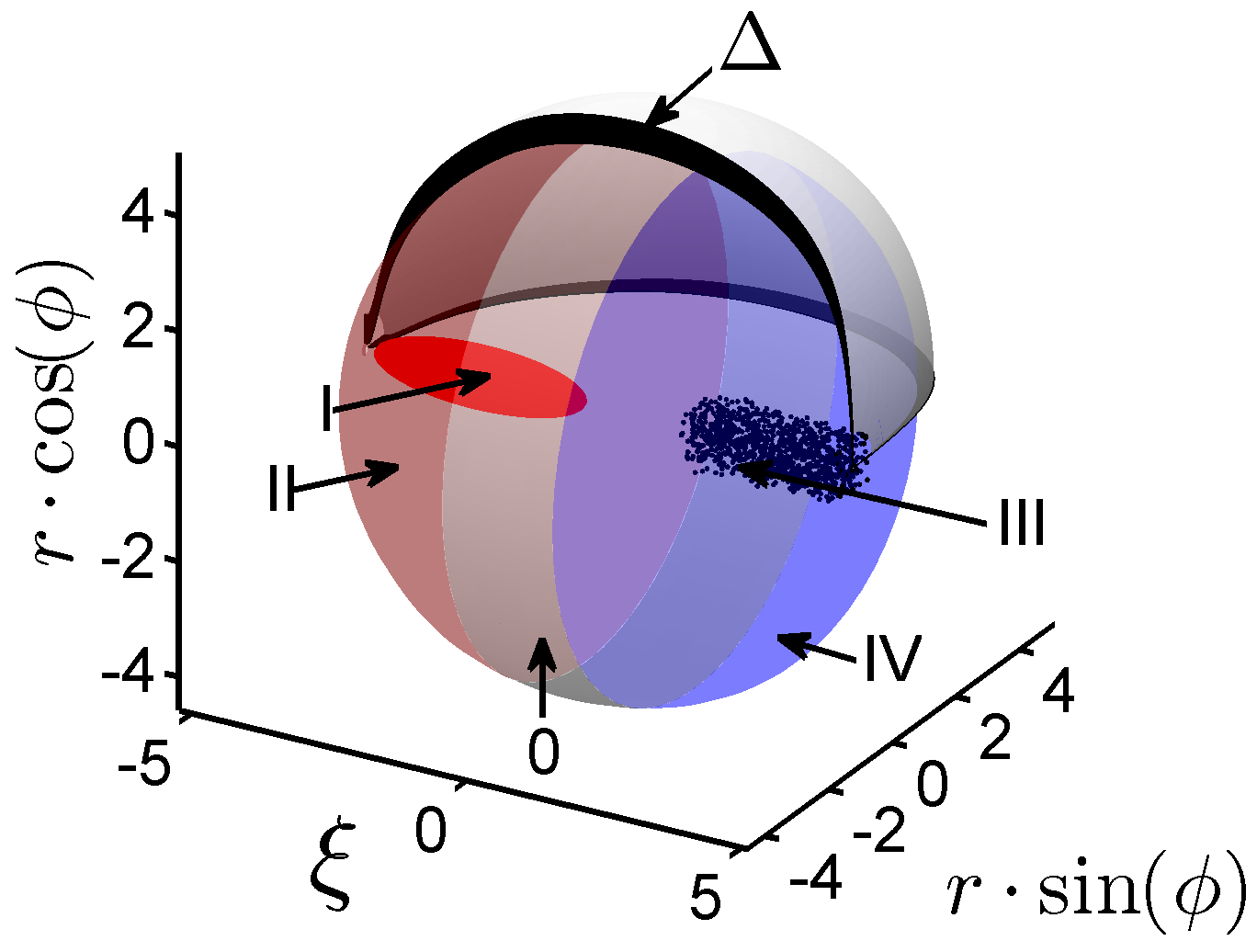}
	\caption{Partition of the five zone blow-out model.  Region 0 (gray): no electron bunch; Region I (red): inside the driving electron bunch; Region II (light red): outside the driving electron bunch; Region III (black dots): inside the accelerated electron bunch; Region IV (light blue): outside the accelerated electron bunch. The inner blow out radius $r_b$ determines the accelerating field. The layer thickness $\Delta$ of the (black) electron layer allows a drop of the high potential at the inner margin to  zero in the free plasma.}
	\label{model}
\end{figure}

To develop the most general model for particle driven blow-outs and laser pulse driven bubbles in a deep plasma channel we first model the normalized electron (current) density inside the ion cavity. Since the cavity is free from background electrons, the only sources we have to consider are the driver and the trapped bunch. Both generate a density $\rho_e$ and a current $J_e$ in propagation direction. In the relativistic case $\rho_e-J_e=\rho_e(1-v_{z,e})\approx 0$ and the electric and magnetic forces from the plasma currents and the beams self-forces each cancel. Thus, in the case of a beam-driven blow-out the electron source can be written as
\begin{align}
& \rho_e = J_e =
\begin{cases}	
j_d(\xi,r),	& r<R_d,\quad \xi_d<\xi<\xi_d+l_d\\
j_b(\xi,r), & r<R_b,\quad \xi_b<\xi<\xi_b+l_b\\
0, 			& \text{else}
\end{cases}.\label{Jedel}
\end{align}
An additional term for the electrons in the blow-out surrounding sheath is modeled later. Trapped electrons perform betatron oscillations and thus generate a part of the radial current $J_r$ each. Nevertheless the average radial current in the electron bunch is zero so that a model of $J_r$ is not necessary. Each limit in Eq.(\ref{Jedel}) - where \textit{d} is for the driver and \textit{b} is for the accelerated bunch - is defined in Fig.\ref{model} while a cylindrical symmetry is assumed. If the driving electron bunch is modeled as a Bi-Gaussian, the limits correspond to $r_e\approx3\sigma_r$ and $l_e\approx3\sigma_\xi$. Since the two bunches generate a current inside the cavity that also acts on the blow-out shape, it is useful to divide the cavity interior into five different zones, shown in Fig.\ref{model}. If the driver is a short laser pulse the first case in Eq.(\ref{Jedel}) does not exist and regions I and II in Fig.\ref{model} merge.

The radially symmetric ion density $\rho_{ion}(r)$ is modeled far simpler because it does not distinguish between the inner blow-out and the outer plasma. The source term for the longitudinal current density $J_z$ and the charge density $\rho$, however, is divided into the three parts
\begin{align}
& S(\xi,r) = J_z-\rho =
\begin{cases}
 s_i(r),\quad r\leq r_b \\
 s_0(\xi),\quad r_b\leq r\leq r_b+\Delta
 \end{cases}\label{s0d}
\end{align}
and $S(\xi,r)=0$ else. Then the source inside the blowout depends only on the distance to the symmetry axis while the source inside the electron layer depends on $\xi$ solely. Since for relativistic electrons in the blow-out regime it is $\rho_e-J_e=\rho_e(1-v_{z,e})\approx 0$ the source inside the blow-out $(r<r_b)$ is just $S=S(r)=s_i(r)=-\rho_{ion}(r)$. Inside the electron sheath an analytical model of the term $\rho-J_z$ is still not known but it was shown that the idea of a source that is constant in $r$ leads to acceptable results \cite{Lu2006}. In our model we also assume that the longitudinal electron current in the sheath can be neglected so that it is dominantly the unknown electron density profile that screens the ion density from the blow-out. The resulting source term $s_0(\xi)$ then should not have any radial dependency left.

In the quasi-static approximation it is $\partial_t\equiv\partial_\xi$, $\partial_z\equiv-\partial_\xi$, and the blow-out potentials are expressed in terms of the vector potential \Ab and the wake field potential $\Psi = \varphi - A_z$. The continuity equation in the cylindrical geometry under the quasi-static approximation $\xi=t-z$ is $r\partial_\xi(\rho-J_z) + \partial_r(rJ_r) = 0$ which links the sheath source to the inner source via 
\begin{align}
s_0(\xi) = \frac{-2c-2S_I(r_b(\xi))}{r_b^2(\xi)[(1+\epsilon)^2 - 1]}. \label{s0}
\end{align}
Here we introduced the integral source 
\begin{align}
S_I(r) = \int_0^{r} s_i(r')r'dr'
\end{align}
and the relative sheath width 
\begin{align}
	\epsilon(\xi) = \frac{\Delta}{r_b(\xi)}.
\end{align}
In the present model we use the Lorenz gauge 
\begin{align}
	\frac{\partial}{\partial r}(rA_r) = -r \frac{\partial}{\partial \xi}\Psi \label{LG}
\end{align}
which gives the normalized Poisson equations
\begin{align}
\frac{1}{r}\frac{\partial}{\partial r}\left(r\frac{\partial A_z}{\partial r}\right) = -J_z, && 	\frac{1}{r}\frac{\partial}{\partial r}\left(r\frac{\partial \Psi}{\partial r}\right) = -\rho + J_z. \label{PG}
\end{align}
From the second equation we find the most general form of the wake field potential
\begin{align}
&\Psi(\xi,r) = \int_0^r\frac{dy}{y}\int_0^yxS(\xi,x)dx + \Psi_0(\xi)
\end{align}
which can also be written in the form $\Psi(\xi,r) = I(\xi,r) + \Psi_0(\xi)$. To determine $I$ and $\Psi_0$ in every region of the blow-out we start with the simplest one - the region of neutral plasma. Here $\Psi=0$ and $r>r_b+\Delta$ so that $\Psi_0=-I$. Since $I$ is the radial integral through all inner regions, we can write $I=I_1+I_2+I_3$ where
\begin{align}
&	I_1(\xi,r) = \int_0^{r_b}\frac{dy}{y}\int_0^yxS(\xi,x)dx = \int_0^{r_b}\frac{S_I(y)}{y}dy\\
&	I_2(\xi,r) = \int_{r_b}^{r_b+\Delta}\frac{dy}{y}\int_0^yxS(\xi,x)dx	\\
&	I_3(\xi,r) = \int_{r_b+\Delta}^r\frac{dy}{y}\int_0^yxS(\xi,x)dx.
\end{align}
With Eq.(\ref{s0}) it follows that $I_3 = -c\int_{r_b+\Delta}^r\frac{dy}{y}$. To determine the integration constant $c$ we claim that the sheath source $s_0$ vanishes at the left and right blow-out limits. Here, the blow-out radius is zero so that $c=0$ is the only possible option. To shorten the explicit expression of $I_2$ we introduce 
\begin{align}
&\beta(\xi) = 2\delta(\xi)\ln(1+\epsilon)-1, \\
&\delta(\xi) = \frac{(1+\epsilon)^2}{(1+\epsilon)^2-1}. \label{delta}
\end{align}
Then it is 
\begin{align}
& I_2(\xi,r) = \int_{r_b}^{r_b+\Delta}\frac{dy}{y}\left(S_I(r_b) + \frac{y^2-r_b^2}{2}s_0(\xi)\right) \nonumber\\
&  = S_I(r_b)\delta(\xi)\ln(1+\epsilon) - \frac{S_I(r_b)}{2} = \frac{S_I(r_b)}{2}\beta(\xi).
\end{align}
This together with $I_1$ gives the value of
\begin{align}
&\Psi_0(\xi) = - \int_0^{r_b}\frac{S_I(y)}{y}dy - \frac{S_I(r_b)}{2}\beta(\xi).
\end{align}
In the interior of the sheath $(r_b < r < r_b+\Delta)$ the integral reduces to $I=I_1+I_4$ where
\begin{align}
& I_4(\xi,r) = \int_{r_b}^{r}\frac{dy}{y}\left(S_I(r_b) + \frac{y^2-r_b^2}{2}s_0(\xi)\right)
\end{align}
is similar to $I_2$. In the blow-out interior the integral reduces to $I=\int_0^rS_I(\xi,y)y^{-1} dy$. Summarizing we write $\Psi$ explicitly as
\begin{align}
\Psi(\xi,r) = \int_0^r\frac{S_I(y)}{y} dy + \Psi_0(\xi), \label{Psi}
\end{align}
for $r \leq r_b$,
\begin{align}
	\Psi(\xi,r) = s_0\frac{r^2}{4} + \frac{S_I(r_b)}{2}\delta\left(2\ln\left(\frac{r}{(1+\epsilon)r_b}\right) +1\right) \label{Psi1}
\end{align}
for $r_b < r < r_b+\Delta$, and $\Psi(\xi,r)=0$ else.

To calculate the longitudinal component of the vector potential first we remind that $A_z$ only depends on the electron current in both the driving and the accelerated bunch $J_e$. Since the bunches are located in the interior, Eq.(\ref{PG}) gives
\begin{align}
r\frac{\partial A_z}{\partial r} = -\int_0^{r} J_z(\xi,r') r'dr'. \label{dazdel}
\end{align}
If we introduce the integral currents $J_D(\xi,r) = \int_0^r j_d(\xi,r') r'dr'$ and $J_B(\xi,r) = \int_0^r j_b(\xi,r') r'dr'$ the solution to $\partial_r A_z$ in all five regions are
\begin{align}
& \frac{\partial A_z}{\partial r} = -\frac{J_D(\xi,r)}{r} &&\text{ in (I)}, \label{eqn:dAzI}\\
& \frac{\partial A_z}{\partial r} = -\frac{J_D(\xi,R_d)}{r} &&\text{ in (II)}	,\\
& \frac{\partial A_z}{\partial r} = -\frac{J_B(\xi,r)}{r}  &&\text{ in (III)} \\
& \frac{\partial A_z}{\partial r} = -\frac{J_B(\xi,R_b)}{r} &&\text{ in (IV)}\label{eqn:dAzIII},
\end{align}
\normalsize
and $\partial_r A_z=0$ else. For the radial component $A_r$ we define
\begin{align}
\sigma(\xi) = -\frac{1}{2}\frac{d \Psi_0(\xi)}{d \xi}.
\end{align}
Then the Lorentz gauge (\ref{LG}) gives
\begin{align}
A_r(\xi,r) = -\frac{1}{2}r\frac{d \Psi_0(\xi)}{d \xi} = r\sigma(\xi)	\label{ARid}
\end{align}
and all components of the electromagnetic potentials inside the cavity are known. In the next section we derive an ODE for the shape of the surrounding electron sheath from these potentials and discuss how the tailored plasma density affects the blow-out shape.

\subsection{The cavity deformation} \label{sheath}
The trajectory of the electrons that form the layer around the cavity are governed by the equations
\begin{align}
\frac{d\pb}{dt} = -\Eb - \frac{\pb}{\gamma}\times\Bb + \Fb, \\ \frac{d\xi}{dt} = 1 - \frac{p_z}{\gamma}, \hspace{1cm} \frac{dr}{dt} = \frac{p_r}{\gamma}\label{r},
\end{align}
where $\pb$ is the kinetic momentum. In the case of the bubble regime the additional force \Fb is the ponderomotive force $\Fb_p=-\gamma^{-1}\nabla|a|^2/4$ of the driving laser pulse \cite{Mora1997}. In the blow-out regime \Fb does not exist. Due to the cylindrical symmetry of our model it is sufficient to describe the dynamics of a test electron with a set of reduced force equations
\begin{align}
\hspace{-.2cm}\frac{dp_r}{dt} = -\left(E_r - \frac{p_z}{\gamma}B_\varphi \right), \frac{dp_z}{dt} = -\left(E_z + \frac{p_r}{\gamma}B_\varphi\right).
\end{align}
The Hamiltonian - here written in terms of the canonical momentum in the moving frame- $H(\Pb,r,\xi,t)=\gamma + P_z - \varphi = \gamma - p_z - \Psi$ is a constant of motion because the present model suggests that the blow-out radius is time-independent. In quiescent plasma $H=1$ so that $d\xi/dt = (1+\Psi)/\gamma$ and $p_r = \gamma\dot{r} = (1+\Psi)dr/d\xi$. To calculate the energy in terms of $p_r$ and $\Psi$, we start with $\gamma^2 = 1 + p_r^2 + p_z^2$ and use the well known relation $\gamma = (1+p_r^2+(1+\Psi)^2)/(2(1+\Psi))$ \cite{Mora1997}, which immediately gives
\begin{align}
& \frac{1}{1-v_z} = \frac{\gamma}{1+\Psi} = \frac{1+p_r^2+(1+\Psi)^2}{2(1+\Psi)^2}.
\end{align}
Now the radial equation of motion can be derived from $dp_r/d\xi$ if the relation between $E_r$, $B_\varphi$ and $\Ab$, $\Psi$ is known. The necessary discussion therefore is given in the next section. Here we only apply the results from Eq.(\ref{EBAP}) and Eq.(\ref{EBAP2}) so that
\begin{multline}
\frac{d\Psi(\xi,r)}{d\xi}\frac{dr}{d\xi} + (1+\Psi)\frac{d^2r}{d\xi^2} =\\
\left[\frac{1+p_r^2+(1+\Psi)^2}{2(1+\Psi)^2}\right]\frac{\partial \Psi(\xi,r)}{\partial r} + \frac{\partial A_z}{\partial r} + \frac{\partial A_r}{\partial \xi} \label{dpsi0id}.
\end{multline}
To determine the electron current at the inner sheath limit we need to solve the equation of motion for a test electron at positions $r=r_b(\xi)$. Here $\Psi(\xi,r_b) = -S_I(r_b)\beta/2$ and the derivatives of the potentials are (also see Eqs. (\ref{eqn:dAzI}) - (\ref{eqn:dAzIII}))
\begin{align}
&\frac{d\Psi}{d\xi} = - \frac{s_i(r_b)r_b}{2}\beta(r_b)r_b' - \frac{S_I(r_b)}{2}\beta'(r_b)r_b', \\
&\frac{\partial\Psi}{\partial r} = \frac{S_I(r_b)}{r_b}, \quad \frac{\partial A_r}{\partial \xi} = r_b\frac{d\sigma}{d\xi},
\end{align}
\begin{align}
\frac{d\sigma}{d\xi} &= \frac{1}{4}\left[2\frac{S_I(r_b)}{r_b} + s_i(r_b)r_b\beta + S_I(r_b)\beta'\right]r_b'' \nonumber\\
& + \frac{1}{4}\left[S_I(r_b)\beta''+ 2s_i(r_b) - 2\frac{S_I(r_b)}{r_b^2}\right](r_b')^2 \label{dsigma}\\
& + \frac{1}{4}\left[s_i(r_b)\beta+ 2s_i(r_b)r_b\beta' + s_i'(r_b)r_b\beta\right](r_b')^2.\nonumber
\end{align}
Here, the derivatives $\beta'$, $\beta''$, $r_b'$, and $r_b''$ are $\beta'(r_b) = d\beta(r_b)/dr_b$, $\beta''(r_b) = d^2\beta(r_b)/dr_b^2$, $r_b'(\xi) = dr_b(\xi)/d\xi$, and $r_b''(\xi) = d^2r_b(\xi)/d\xi^2$. If we sort Eq.(\ref{dpsi0id}) with respect to derivations of $r_b$ and substitute the expressions above we find
\begin{align}
A(r_b)r_b'' + B(r_b)(r_b')^2 + C(r_b) = \frac{\Lambda(\xi)}{r_b} \label{ODE}
\end{align}
The coefficient functions are
\begin{align}
	& A(r_b) = 1 - \frac{S_I(r_b)}{2} - \frac{1}{4}(2S_I(r_b) + s_i(r_b)r_b^2)\beta \nonumber\\
	& \hspace{1.5cm} - \frac{r_b}{4}S_I(r_b)\beta', \\
	& B(r_b) =  -\frac{s_i(r_b)r_b}{2} - \left[3s_i(r_b)r_b + s_i'(r_b)r_b^2\right]\frac{\beta}{4} \nonumber\\
	& \hspace{1.5cm} - [S_I(r_b)+s_i(r_b)r_b^2]\frac{\beta'}{2} - S_I(r_b)r_b\frac{\beta''}{4},\\ 
	& C(r_b) = -\frac{S_I(r_b)}{2r_b}\left(1 + \left(1- \frac{S_I(r_b)}{2}\beta\right)^{-2}\right),
\end{align}
and $\Lambda(\xi)=-J_D(\xi,R_d)$ at the driving bunch, $\Lambda(\xi)=-J_B(\xi,R_b)$ at the driven bunch, as well as $\Lambda(\xi)=0$ else. If a laser pulse drives a bubble we have to consider the ponderomotive force $\Fb_p$ again so that, according to \cite{Mora1997}, 
$C(r_b)= -\frac{S_I(r_b)}{2r_b}\left(1 + \left(1+\frac{|a|^2}{2}\right)\left(1- \frac{S_I(r_b)}{2}\beta\right)^{-2}\right)$ and the right hand side of Eq.(\ref{ODE}) becomes $-d|a|^2/dr/(4-2S_I(r_b)\beta)$.

In the special case of a constant ion density these coefficients reduce to the well known ones from Lu et al. \cite{Lu2006}. Here a discussion of the limit $\epsilon\rightarrow0$ showed that a blowout can be approximated by a perfect sphere if the distance of the border to driver axis is sufficiently large. In all other cases, the shape of the electron layer is flattened. In our generalized model we observe the same matter except that we can give a more general connection between the plasma channel shape and the layer form. For example, if we assume a sufficient large potential at the inner border (which can be found at the point of maximal $r_b$) the ODE coefficients simplify to
\begin{align}
& A(r_b) = 1 - \frac{S_I(r_b)}{2}, && B(r_b) = - \frac{s_i(r_b)r_b}{2}, \label{AB}\\ & C(r_b) = -\frac{S_I(r_b)}{2r_b}. \label{C}
\end{align}
In the further special case of a homogeneous plasma these functions further reduce to $A=1+r_b^2/4$, $B=r_b/2$, and $C=r_b/4$, which in turn leads to the ODE
\begin{align}
 & \left(1+\frac{r_b^2}{4}\right)r_br_b'' + \frac{r_b^2}{2}(r_b')^2 + \frac{r_b^2}{4} = \Lambda.
\end{align}
The solution for large $r_b$ and without considering the action of the driver or beam load resembles a circle with radius $R$, centered on the driver axis. This is why former analytical models assumed that the bubble is a perfect sphere \cite{Kostyukov2004, Kostyukov2009, Kostyukov2010, Kalmykov2011, Kalmykov2012}.

\begin{figure}[t]
	\centering
	\includegraphics[width=0.95\columnwidth]{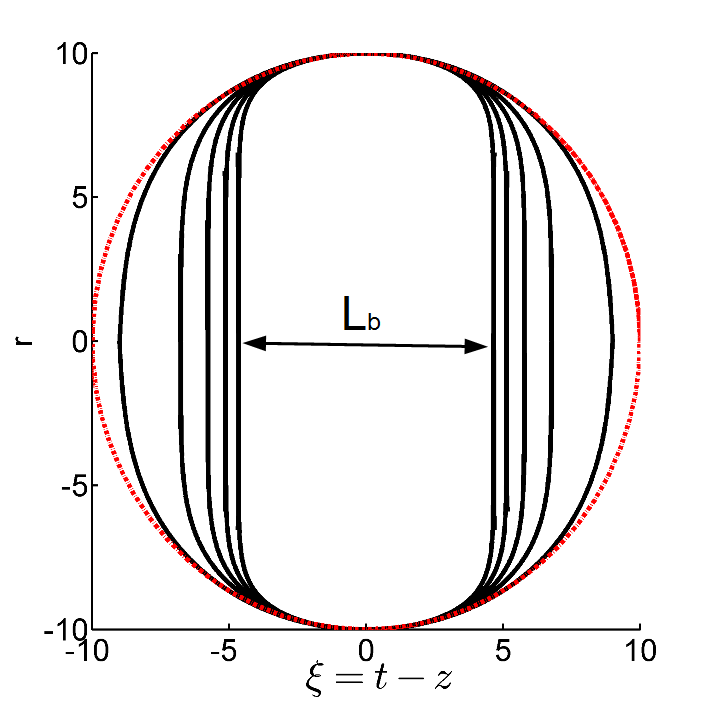}
	\caption{Electron layer shapes for plasma channels (black solid lines) of the form $\rho_{ion}=r^k$, $k=0,2,4,6,8$ and former spherical profile (red dotted line) from homogeneous plasma models.}
	\label{shape}
\end{figure}

For a more general plasma density profile of the form $\rho_{ion}=r^k$ the electron layer shape steepens at the front and the rear with increasing $k$. This effect is shown in Fig.\ref{shape} where the red dotted circle represents the sphere that usually serves as a model for the bubble. The black flattened circles are solutions of Eq.(\ref{ODE}) for $k=0,2,4,6,8$ and $\epsilon\rightarrow0$.

To explain why a deep plasma channel leads to a squeezed blow-out shape we restrict to a general polynomial plasma channel of the form $\rho_{ion}(r)=\alpha r^n$ with $n\geq 0$ and $\alpha>0$. Then, according to Eq.(\ref{ODE}) the ODE for the shape reduces in the limit $\alpha r_{b}^{n+2}/(2n+4)\gg1$ and for a neglected bunch action to 
\begin{align}
	r_{b}\frac{d^{2}r_{b}}{d\xi^{2}}+(n+2)\left(\frac{dr_{b}}{d\xi}\right)^{2}+1=0.
\end{align}
This equation describes a flattened circle - not an ellipsis. The analytical solution can be expressed implicitly in terms of the blow-out radius
\begin{align}
	\xi=\frac{r_b^{n+3}}{R^{n+2}}\frac{\sqrt{n+2}}{n+3}F_{2,1}\left(\frac{n+3}{2n+4},\frac{1}{2};\frac{3n+7}{2n+4};\frac{r_{b}^{2n+4}}{R^{2n+4}}\right).
\end{align}
Here $R=\max r_b$ is the maximum blow-out radius and $r_b(\xi=0)=0$.  The Hypergeometric function 
\begin{align}
F_{2,1}(a_1,a_2;b;z)=\sum_{k=0}^\infty\prod_{i=1}^2\frac{\Gamma(k+a_i)}{\Gamma(a_i)}\frac{\Gamma(b)}{\Gamma(k+b)}\frac{z^k}{k!}
\end{align}
can be evaluated explicitly if we set $r_b=R$. So it is possible to derive the blow-out length in the simple form
\begin{align}
	L_{b}=2\xi\left(r_{b}=R\right) = 2R\sqrt{\pi\left(n+2\right)}\frac{\Gamma\left(\frac{n+3}{2n+4}\right)}{\Gamma\left(\frac{1}{2n+4}\right)}.
\end{align}
This function indeed decreases with increasing $n$. If the action of the electron bunches on the sheath is taken into account, the shape might differ from this again. In general, however, we expect that the blow-out gets compacter and that its front and rear get steeper.
\begin{figure}[t]
	\centering
	\includegraphics[width=1\columnwidth]{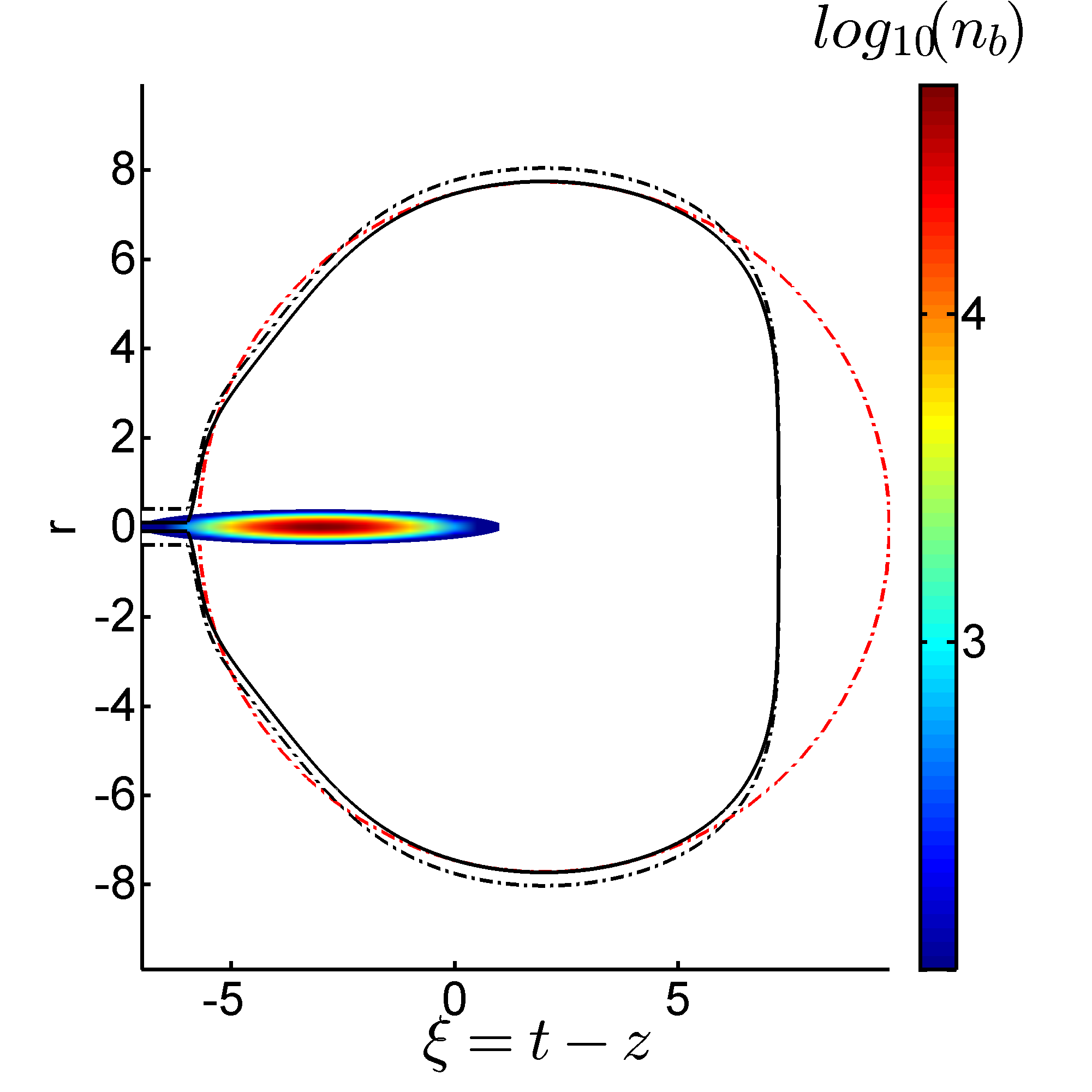}
	\caption{Deformation of the blow-out shape (black line) with electron layer (black dotted line) in a parabolic plasma channel driven by an electron bunch with $\sigma_z=1$, $\sigma_r=0.1$, and $n_{b,0}=-5\text{x}10^4$. The former spherical profile is added as a red dotted circle.}
	\label{layer}
\end{figure}

In PIC simulations the squeezing effect is not that obvious because the driver and the beam load act back on the sheath. In Fig.\ref{layer} this difference is demonstrated for $\rho_{ion}=r^2$ and a bi-Gaussian electron driver with $k_p\sigma_z=1$ and $k_p\sigma_r=0.1$. In the figure the layer in front of the driver is as steep as at the blow-out rear while the sheath is pushed over the limit of the previously modeled spherical blow-out shape. Due to this elongation the electron cavity follows the circular form for a longer distance. In the back the above described shortening is obvious again.

\begin{figure}[t]
	\includegraphics[width=1\columnwidth]{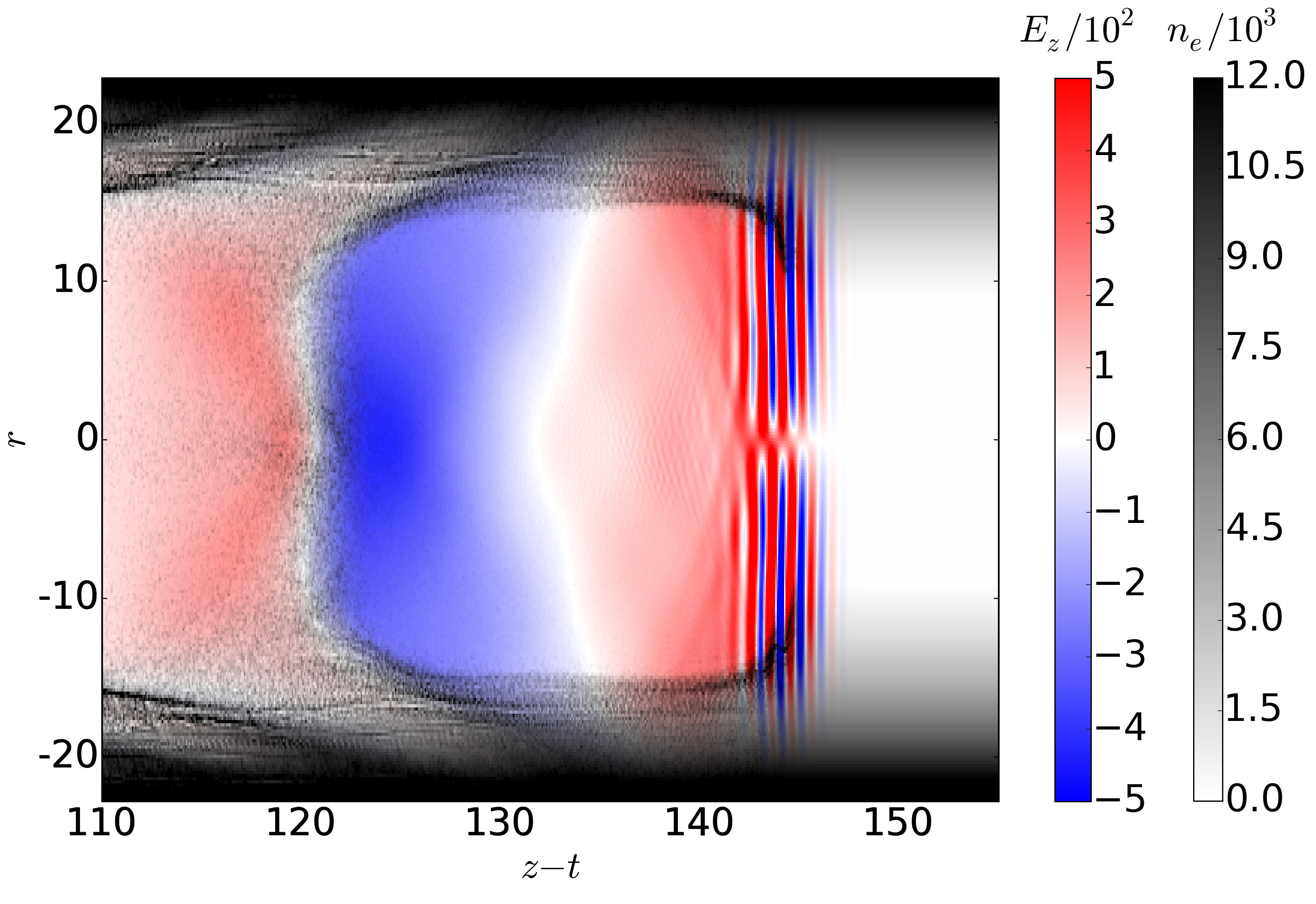}
	\caption{\label{fig:Exn}  Longitudinal bubble field $E_{z}$ and cross-section of the plasma electron density from a PIC simulation. The driving laser pulse has focal spot size $R=8\,\mu$m and $a_0=10$. The purple rippling of $E_{z}$ at the bubble head is caused by the short laser pulse.}
\end{figure}

In the laser-driven case the front side is not pushed over the analytical limit, which is demonstrated in Fig.\ref{fig:Exn}. Here we applied a plasma density that contains a vacuum part around the symmetry axis. We have chosen the following parametrization for the plasma radial profile: $n_{e}=n_{0}\left[\delta_{ch}\exp(r/R_{{\rm ch}})-1\right]$, for\emph{ $r\geq r_{0}$ }\textsl{\emph{and }}\emph{$n_{e}=0$ } \textsl{\emph{for} }\emph{$r<r_{0}$, }\textsl{\emph{where} }\emph{$r_{0}=R_{ch} \ln(1/\delta_{ch})$}. The radially exponential density profile suggests that the channel is produced by ablation from walls of an empty capillary, e.g. heated by a prepulse \cite{Katzir2009}. In our simulations we used a single highly intense $800\,$ nm laser pulse with focal spot size $R=10\,\lambda_{laser}$, duration $\tau=4\,$ fs, $a_{0}=10$, and an underdense ambient plasma with $n_0=0.0029n_c$, where $n_c$ is the critical density of the plasma. The main reason why the laser pulse does not push the bubble shape far in propagation direction is that it is very short. Thus the ponderomotive force edges the bubble at the front but does not act long enough to elongate its sheath well along the analytically predicted limit.

In the next section we calculate the fields from the potentials we have found both in the blow-out interior and inside the sheath. Afterwards we discuss in how far the sources can be chosen to generate a given field configuration.

\subsection{Fields and sources} \label{field}
Now that we have found the potentials, the equations of motion for test electrons, and an equation for the electron sheath current, we are ready to calculate the fields in terms of the sources and vice versa. Since we have different solutions to $\partial_r A_z$ - one for each of five zones inside the blow-out - we will focus on a case sensitive solution for the fields, too. For the fields it is possible to give straightforward expressions in the case that neither the accelerated nor the driving electron bunch affect the sheath. Now, however, we want to include all cases into the found expressions. 

\subsubsection{Fields from sources}
From our model of the new sources we have $\rho = \rho_e + \rho_{ion} = j_{d,b}(\xi,r) + \rho_{ion}$, $\Jb = J_z\vec{e}_z = (s_i(r)+\rho)\vec{e}_z = (j_{d,b}(\xi,r) + s_i(r) + \rho_{ion})\vec{e}_z$, and $J_i = 0$. The cylindrical symmetry and the quasi-static approximation give the general form of the electromagnetic fields
\begin{align}
	E_z & = \frac{\partial \Psi}{\partial \xi},	\quad B_\varphi = - \frac{\partial A_r}{\partial \xi} - \frac{\partial A_z}{\partial r}, \label{EBAP}\\
	E_r & = -\frac{\partial \Psi}{\partial r} - \frac{\partial A_r}{\partial \xi} - \frac{\partial A_z}{\partial r}.\label{EBAP2}
\end{align}
For $r\leq r_b$ this and Eq.(\ref{Psi}) give the fields inside the blow-out in a plasma channel
\begin{align}
&  E_z = -2\sigma, \quad B_\varphi = -r\frac{d\sigma}{d\xi} - \frac{\partial A_z}{\partial r}, \label{Ezin}\\
& E_r = -\frac{S_I(r)}{r} - r\frac{d\sigma}{d\xi} - \frac{\partial A_z}{\partial r}. \label{Erin}
\end{align}
In the special case of an empty, spherical blow-out without an electron sheath in a homogeneous plasma these components reduce to the well known functions $E_z=-\xi/2$, $E_r=r/4$, and $B_\varphi=-r/4$ \cite{Kostyukov2004,Kostyukov2009}.

The fields inside the electron layer are rather complicated as $\Psi$ and its derivatives strongly depend on $r$ and $\xi$ in this region. The latter can be expressed simplified for arbitrary sheath thicknesses if we introduce $X=1+\epsilon$, $s_1(\xi) = S_I(r_b) \delta(\xi)$ (cmp. Eq.(\ref{delta})), and 
\begin{align}
	\Psi_1(\xi) & = - \frac{S_I(r_b)}{2} \delta (2 \ln(r_b X) -1).
\end{align}
In terms of the abbreviations $s'_{ib}=s_i'(r_b)$, $s_{ib}=s_i(r_b)$ and $S_{Ib}=S_I(r_b)$ the longitudinal electrical field inside the sheath is
\begin{align}
	&E_z = \frac{\partial}{\partial \xi} \left( \frac{1}{4}s_0(\xi)r^2 + s_1(\xi)\ln(r) + \Psi_1(\xi) \right)
	\nonumber\\
	&=  - \frac{1}{2}r^2 \frac{s_{ib} + \epsilon s_0(\xi)}{X^2-1}\frac{r_b'}{r_b} + \frac{s_{ib}\delta r_br_b'}{2}  - \frac{S_{Ib} \delta}{X+1} \frac{r_b'}{r_b}
	\nonumber
	\\ 
	 &\hspace{.35cm}+ \ln\left(\frac{r}{r_bX}\right)\delta r_b'\left(s_{ib}r_b + \frac{2S_{Ib}}{r_bX(X+1)}\right)
\end{align}
while the radial electric field is a composition of 
\begin{align}
	\frac{\partial \Psi}{\partial r}(\xi,r) & = \frac{\partial}{\partial r} \left(\frac{s_0(\xi)}{4}r^2 + s_1(\xi)\ln(r) + \Psi_1(\xi) \right)
	\nonumber\\
	& = \frac{1}{2} s_0(\xi) r + \frac{1}{r} s_1(\xi)
\end{align}
and $B_\varphi$. To determine the latter one we could use the Lorentz gauge (\ref{LG}) and the normalized Poisson equations again. Unfortunately, the integral term that must be computed can not be simplified like Eq.(\ref{ARid}) because $\partial\Psi/\partial \xi$ now explicitly depends on $r$. However, we want to find a closed expression for $E_r$ and $B_\varphi$. Thus we assume again that $\epsilon$ is a smallness parameter for large $r_b$ and approximate
\begin{align}
	\delta & = \frac{1 + 2 \epsilon + \epsilon^2}{2 \epsilon + \epsilon ^2} \approx \frac{\epsilon}{8} + \frac{3}{4} + \frac{1}{2\epsilon},\\
	s_0 & =  - \frac{2S_{Ib}}{r_b^2(2 \epsilon + \epsilon^2)} \approx -\frac{S_{Ib}}{r_b^2\epsilon}\left(1-\frac{\epsilon}{2}+\frac{\epsilon^2}{4}\right), \label{s0a} \\
	 s_1 & =  S_{Ib} \delta \approx S_{Ib}\left(\frac{\epsilon}{8}+\frac{3}{4} + \frac{1}{2\epsilon}\right).
\end{align}
Within this approach we find a second smallness parameter giving the position inside the electron layer relative to the blow out radius $\zeta=(r-r_b)/r_b$ so that $0\leq\zeta\leq\epsilon$. If we substitute $\zeta$ into the wakefield and linearize the logarithmic parts most parts cancel out and, altogether, the wake field potential inside the electron sheath simplifies to
\begin{align}
	\Psi(\xi,r) \approx -\frac{S_{Ib}}{2\epsilon}(\zeta - \epsilon)^2.
\end{align}
This potential indeed fulfills $\Psi(r=r_b)=-\epsilon S_{Ib}/2$ and vanishes for $r = r_b+\Delta$. It is quadratic in $r$ so that its contribution to the radial electrical field
\begin{align}
\frac{\partial \Psi}{\partial r}(\xi,r) = \frac{(\epsilon-\zeta)}{\epsilon}\frac{S_{Ib}}{r_b} \label{dpsir}
\end{align}
is linear in $r$. The derivatives of $\Psi$ with respect to $\xi$ must be treated a bit more cautious because $\partial\zeta/\partial\xi=-(\zeta+1)r_b'/r_b$ which means that higher order terms in $\Psi$ contribute to lower terms in the derivative. Fortunately, from all higher terms only the second order terms (which we have neglected) in $\Psi$ contribute to $E_z$ so that in the limit of our precision
\begin{align}
E_z  &\approx -\frac{(\epsilon-\zeta)}{\epsilon}S_{Ib}\frac{r_b'}{r_b}.
\end{align}
For the magnetic field inside the blow-out $(r\leq r_b)$ we found short abbreviations like $\partial A_r/\partial\xi=r\sigma'$ and $\partial A_z/\partial r=\Lambda/r$. For the fields within the electron layer $(r>r_b)$ a direct computation of 
\begin{align}
\frac{\partial A_r}{\partial \xi}  = - \frac{1}{r} \int_0^r \frac{\partial^2 \Psi(\xi,r')}{\partial^2\xi} r' dr'
\end{align}
and 
\begin{align}
\frac{\partial A_z}{\partial r} = -\frac{1}{r}\int_0^{r} J_z(\xi,r') r'dr'
\end{align}
is necessary, while the second part needs knowledge about the currents inside the layer. These are the plasma return currents which are constant in radial direction. We denote them as $J_s(\xi)$. Since we assumed that the source term $s_0$ for the electron sheath depends only on $\xi$ it is convenient to do the same for the current which is related to the wake field potential via Ampere's circuit law
\begin{align}
\oint_{\partial P}\vec{B}\cdot d\vec{l} = \int_P \left(\vec{J}+\frac{d\vec{E}}{dt}\right)\cdot d\vec{s}.
\end{align}
for our purpose we chose $P$ to be a transverse plane with unit normal vector $\hat{n}=\vec{e}_z$. If we further integrate over the complete plane the boundary vanishes and
\begin{align}
\int_{\mathbb{R}^2} \left(J_z+\frac{\partial E_z}{\partial t}\right) dxdy = 0
\end{align}
which immediately gives
\begin{align}
\int_0^{r_b+\Delta} J_z rdr = -\int_0^{r_b+\Delta}\frac{\partial^2 \Psi}{\partial\xi^2} rdr
\end{align}
because both $\Psi$ and $J_z$ vanish outside the electron sheath. Since $\int_0^{r_b} J_z rdr = -\Lambda(\xi)$ the sheath return current is connected to the wake field potential and the source current by
\begin{align}
\frac{(r_b+\Delta)^2-r_b^2}{2}J_s(\xi) =-\int_0^{r_b+\Delta}\frac{\partial^2 \Psi}{\partial\xi^2} r'dr' + \Lambda(\xi).
\end{align}
Besides this circumstance another important fact is that the magnetic field - and thus also the radial electric field - outside the electron layer $(r>r_b+\Delta)$ completely dissolve. To see this, we combine the Lorentz gauge from Eq.(\ref{LG}) and the normalized Poisson equation (\ref{PG}) for $(r>r_b+\Delta)$ and get
\begin{align}
rB_\varphi = \frac{\partial}{\partial \xi} \left( \int_0^{r_b+\Delta} \frac{\partial \Psi}{\partial \xi} r' dr' \right) - \int_0^{r_b+\Delta}\frac{\partial^2 \Psi}{\partial\xi^2} r' dr' = 0.
\end{align}
For the case $(r_b+\Delta>r>r_b)$ the first integral has upper bound $r$ and the  magnetic field is completely determined by $\Lambda(\xi)$ and $\int\frac{\partial^2 \Psi(\xi,r)}{\partial^2\xi} rdr$ as
\begin{align}
rB_\varphi = \frac{X^2-(\zeta+1)^2}{X^2-1} \left(\int_0^{r_b}\frac{\partial^2 \Psi}{\partial\xi^2} r' dr' -\Lambda(\xi)\right) \nonumber\\- \frac{(\zeta+1)^2-1}{X^2-1}\int_{r_b}^{r_b+\Delta}\frac{\partial^2 \Psi}{\partial\xi^2} r'dr' + \int_{r_b}^r\frac{\partial^2 \Psi}{\partial\xi^2} r'dr', \label{rBr}
\end{align}
where the term in brackets is simply $r_bB_\varphi(r_b)$. According to our level of precision
\begin{align}
	\frac{\partial^2 \Psi}{\partial\xi^2} = \frac{\partial E_z}{\partial\xi}  \approx \frac{S_{Ib}}{\epsilon}\frac{r_b'^2}{r_b^2}, && \frac{X^2-(\zeta+1)^2}{X^2-1}\approx\frac{\epsilon-\zeta}{\epsilon}
\end{align}
and $r_b/r\approx1$ so that the 2nd and 3rd summand in Eq.(\ref{rBr}) cancel out. What remains is the monotone falling 
\begin{align}
	B_\varphi = \frac{(\epsilon-\zeta)}{\epsilon}B_\varphi(r_b)
\end{align}
which gives - together with Eq.(\ref{dpsir}) - the radial electrical field inside the layer
\begin{align}
E_r  = \frac{(\epsilon-\zeta)}{\epsilon}\left(B_\varphi(r_b) - \frac{S_I(r_b)}{r_b}\right).
\end{align}
This set of fields describes the motion of an electron in the blow-out surrounding layer in the same way as the fields in Eqs.(\ref{Ezin}) and (\ref{Erin}) describe the motion inside the blow-out if the inner sheath radius is determined from the ODE (\ref{ODE}) with coefficient function from Eq.(\ref{AB}),(\ref{C}). In this case the inner fields simplify substantially, because
\begin{align}
\sigma \approx \frac{S_{Ib}}{2}\frac{r_b'}{r_b}.
\end{align}

\subsubsection{Sources from reversed fields}
Now we ask under which conditions it is possible to recalculate the sources for given fields. Since we already know the general structure of all possible fields we assume that $B_r=B_z=E_\varphi=0$ and that $B_\varphi$, $E_r$, $E_z$ as well as the solution to $r_b(\xi)$ from the ODE are given. Then it is $E_z=-2\sigma=d\Psi_0/d\xi$ which leads to a first order ODE for $s_i(r_b)$. To avoid this complication we assume again that the electron sheath thickness is much smaller than the inner blow out radius. Then it is $\beta\approx0$ and we can apply Eq.(\ref{s0a}) so that $E_z\approx-S_I(r_b)r_b'/r_b=s_0(\xi)\Delta r_b'$ and thus
\begin{align}
s_0(\xi) = \frac{E_z(\xi)}{\Delta r_b'(\xi)}. \label{s00}
\end{align}
The source for the inner region is
\begin{align}
s_i(r) = \left(\frac{1}{r} + \frac{\partial}{\partial r}\right)(B_\varphi(\xi,r) - E_r(\xi,r)) \label{si}
\end{align}
while the electron current in the first region is always zero and so is its source. In all other regions the source can be deduced from $\partial_r A_z$ and so from $\sigma$ and $B_\varphi$ via:
\begin{align}
& j_d(\xi,r), j_b(\xi,r) = -\left(\frac{1}{r} + \frac{\partial}{\partial r}\right)\left(B_\varphi + r\frac{d\sigma(\xi)}{d\xi}\right)
\end{align}
depending on whether $B$ is taken from region I or III. Since $s_i$ and $s_0$ are related to each other, the choice of $E_z$,  $B_\varphi$, and $E_r$ is not arbitrary. However - once a set of field components is found that self-consistently fulfills Eqs.(\ref{s00}),(\ref{si}), and (\ref{s0}), the corresponding plasma channel model is known too.

In the next section we discuss how the radial density profile can be used as a new degree of freedom to reverse at least one field component and how this can be used to neutralize the focusing force in the blow-out regime.

\subsection{Field reversal in vacuum region}\label{reverse}
\begin{figure}[t]
	\centering
	\includegraphics[width=1\columnwidth]{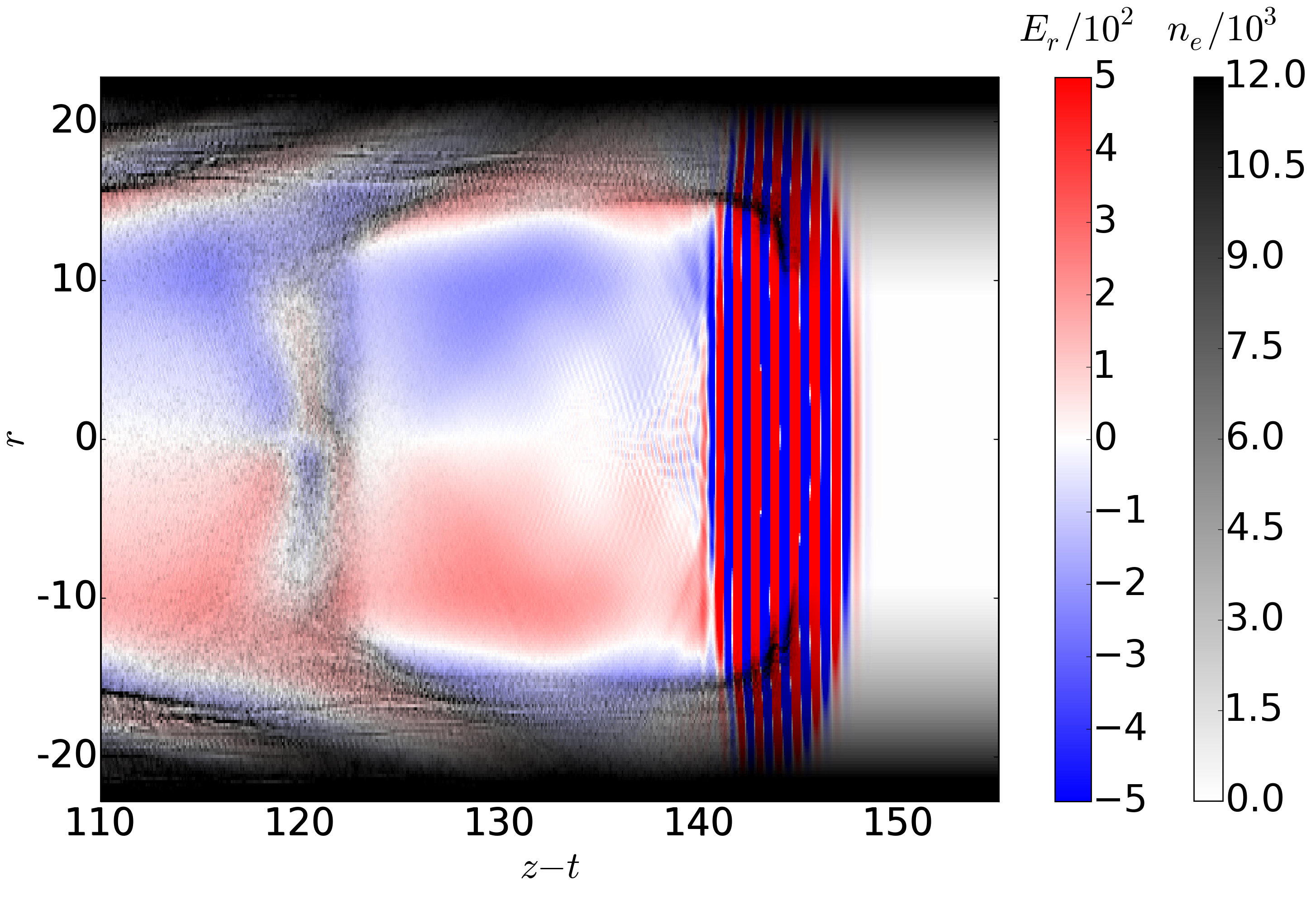}
	\caption{\label{fig:Er} Radial electrical field $E_r$  inside a bubble which is driven by a laser pulse with focal spot size $R=8\,\mu$m. The field has the same sign inside the vacuum part and the beginning of the channel wall.}
\end{figure}
In the last section we showed how the electromagnetic fields inside an electron beam driven blow-out and inside the surrounding electron sheath depend on the plasma density. We also argued that the inverted relations can be used to create (in a certain limit) arbitrary field configurations. However, practically it is quite hard to find a density configuration that fits the desired fields. In particular we could ask whether it is possible to find a plasma density such that the fields inside the blow-out are completely reversed in comparison to the case of a homogeneous plasma - and that without just reversing the sign of the charges. If we do this we conclude that all $\sigma$, $d\sigma/d\xi$, and $S_I(r)$ have to change their sign. This, however, is only possible if we introduced negative densities or reversed the charge sign of the plasma particles. 

Another - weaker - postulate is to change the sign of just one field component. Here, it is convenient to choose $E_r$ because in absence of the driver and beam load it consists of two terms that potentially cancel each other in a deep plasma channel with a vacuum part around the driver axis. From PIC simulations (see Fig.\ref{fig:Er}) we know that the radial electrical field inside the cavity has the same sign independent from whether it is measured in the vacuum part of the channel or in its walls. As  Fig.\ref{fig:Er} shows $E_r$ seems to change its sign in the electron layer. However this effect is rather related to the increasing plasma density which compensates the $r\sigma'$-term in Eq.(\ref{Erin}). As a consequence we further restrict our goal and try to find a plasma channel that leads to a switch of the sign in $E_r$ at any distance from the $\xi$-axis inside the cavity. To start we think about a situation in which $\sgn(E_r(s_i(r))) = -\sgn(E_r(s_i=-1))$ inside the electron bunch free region. In the approximation that the electron layer shape is a sphere with constant radius this is equivalent to
\begin{align}
	& \sgn\left(\frac{S_I(r)}{r} + r\frac{d\sigma}{d\xi}\right) = -\sgn\left(-\frac{r}{2} + \frac{r}{4}\right) = 1.
\end{align}
Since $S_I(r)$ is always negative or zero, its absolute value must be smaller than $rd\sigma/d\xi$. An example that definitely fulfills this condition is a plasma density that drops to zero on axis and around. Then $S_I(r)=0$ in the vacuum region and $rd\sigma/d\xi>0$  if the blow-out radius is larger than the vacuum limit. A practical use of a reversed radial field is the weakening of the focusing force in the blow-out. Inside the vacuum part it is even possible that this force becomes defocussing as $E_r = B_\varphi = -rd\sigma/d\xi - \partial A_z/\partial r$ and thus $F_r=B_\varphi p_z/\gamma -E_r\approx0$. 

\begin{figure}[t]
	\centering
	\includegraphics[width=1\columnwidth]{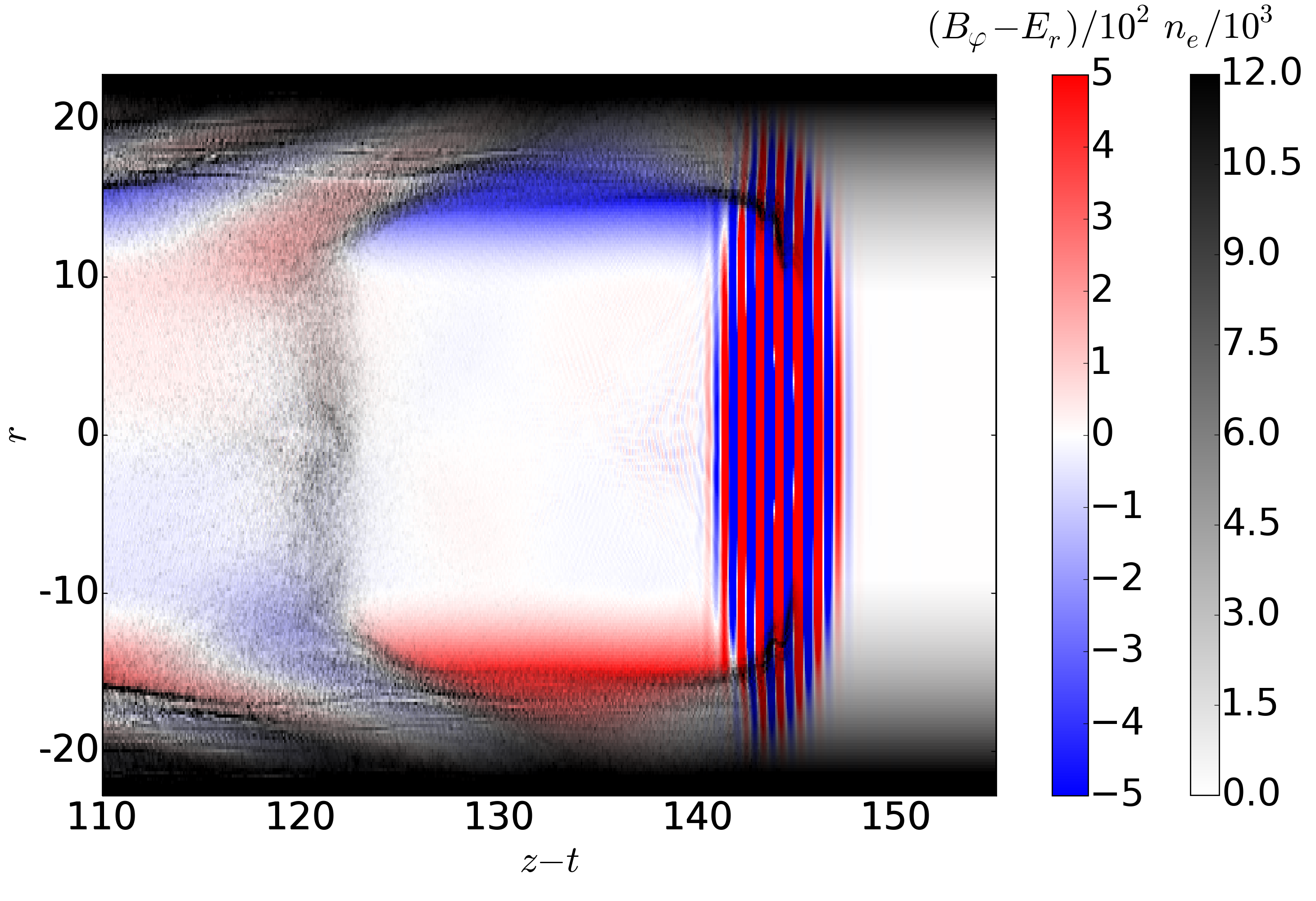}
	\caption{\label{fig:reversal} Focusing force $F_r\approx B_\varphi-E_r$  inside a bubble which is driven by a laser pulse with focal spot size $R=8\,\mu$m and $a_{0}=10$. The force almost completely vanishes inside the vacuum part. What remains inside the walls acts like a containment coat for trapped electrons.}
\end{figure}

To demonstrate the almost complete missing of the focusing force in the vacuum part numerically we calculated $B_\varphi-E_r$ from a PIC simulation for an exponential plasma channel. We chose the plasma radial profile $n_{e}=n_{0}\left[\delta_{ch}\exp(r/R_{{\rm ch}})-1\right]$, for\emph{ $r\geq r_{0}$ }\textsl{\emph{and }}\emph{$n_{e}=0$ } \textsl{\emph{for} }\emph{$r<r_{0}$ } \textsl{\emph{where} }\emph{$r_{0}=R_{ch} \ln(1/\delta_{ch})$} again. The result is shown in Fig.\ref{fig:reversal}. Here we clearly see that the focusing force inside the bubble is much weaker in the vacuum part than in the plasma wall.

The reversal of the radial electrical field is shown in Fig.\ref{fig:Er}. Here we see how the field first vanishes and then changes its sign inside the plasma wall near the bubble border.
Another feature of the reversed radial electrical field is the strongly reduced amplitude. This reduction also follows from the model as the usually dominant integral source term is zero in the vacuum part.

The new parameters $n_{0},\,\delta_{ch}$, and $R_{{\rm ch}}$ provide enough freedom to improve the quality of the accelerated electron bunch and to ensure a smooth guiding of the relativistic laser pulse. Both together can be used to avoid several disadvantages the bubble regime exhibits in uniform plasma. The first and maybe most important disadvantage is that the laser energy depletion length is shorter than the electron dephasing length. This limits the maximum electron energy gain. Second, the betatron eigenfrequency of the accelerated electrons can easily come into resonance with the Doppler-shifted laser wave. This leads to an energy exchange and a broadening of the electron bunch energy distribution which in turn deteriorates the beam quality \cite{Pukhov1999a, Cipiccia2011, Shaw2014a}.

To overcome the second problem it is necessary to reduce the strongly focusing transverse bubble fields. Then, electrons running forward with the relativistic factor $\gamma$ would not oscillate about the bubble axis at the betatron frequency $\omega_{\beta}=\omega_{p}/\sqrt{2\gamma}$. A reversed transversal electric field, as introduced above, offers a simple solution to this problem because it can lead to a complete cancellation of the focusing force. A solution to the first problem of different depletion and dephasing lengths can only be found if we use the new degrees of freedom from the plasma channel. In our former work \cite{Pukhov2014b} this was demonstrated for the above plasma channel. However, the adaption of both limits led to a new optimal scaling of the laser pulse for the bubble regime.

\subsection{Conclusion}
We have introduced an analytical model to describe wake fields in the highly non-linear broken wave regime which are generated by relativistic electron beams or intense short laser pulses propagating through a preformed plasma channel. These wake fields are generated when all electrons are ejected from the driving axis and form a sheath. This sheath is modeled as a finite-size electron layer with a thickness that is negligible compared to the blow-out or bubble radius $r_b$. Our theory describes the equation of motion of an electron in the layer which can be solved for polynomial plasma profiles. The solutions then predict - in agreement with numerical simulations - a shortening and steepening of the cavity profile for steeper plasma channels. The electric and magnetic fields inside the electron cavity are expressed in terms of the background ion density and the electron (current) density in the sheath. The fields in the sheath are calculated as well. A reversal of the transversal electric field and a combined suppression of the focusing force are predicted for an empty channel and confirmed by PIC simulations.
\\

This work has been supported in parts by the Deutsche Forschungsgemeinschaft via SFB TR 18, EU FP7 project EUCARD-2, the Government of the Russian Federation (Project No. 14.B25.31.0008), the Russian Foundation for Basic Research (Grants No. 13-02-00886, 13-02-97025)


\end{document}